\documentclass[prb,preprint,aps]{revtex4-1}
\usepackage{graphicx}
\usepackage{amsmath}
\usepackage{amssymb}
\usepackage{epsfig}
\usepackage{color}

\begin{document}

\title{On the motion of classical three-body system with consideration of
quantum fluctuations}

\author{ Ashot Gevorkyan$^{1,2}$}

\address{$^{1}$ Institute for Informatics and Automation Problems, NAS of RA}
 \address{$^{2}$ Institute of Chemical Physics, NAS of RA, \\
 g\_\,ashot@sci.am}

\begin{abstract}
We obtained the system of stochastic differential equations which describes the
classical motion of the three-body system under influence of quantum fluctuations.
Using  SDEs, for the joint probability distribution of the total momentum of bodies
system  were obtained the  partial differential equation of the second order.
It is shown, that the equation for the probability distribution is solved jointly
by classical equations, which in turn are responsible for the topological peculiarities
of tubes of quantum currents, transitions between asymptotic channels and, respectively
for  arising of quantum chaos.
 \end{abstract}
\pacs{45.50.Jf, 34.50.-s,02.50.Ey,05.45.Mt}

\maketitle

\section{Introduction}
The behavior of quantum systems, non-integrable in the classical limit was discussed
still at the dawn of development of the quantum mechanics \cite{Einst}.
The occurrence of wave mechanics has allowed us to formulate the time evolution of any
quantum system in the framework of the time-dependent Schr\"{o}dinger equation.
Despite the fact that in quantum mechanics essentially possible only a statistical description
due to the linearity of the Schr\"{o}dinger equation, the wave function of a quantum system
has a  deterministic behavior from the time. Last circumstance creates serious difficulties
for the description of the dynamical system at transition to the classical limit where
the system demonstrates a chaotic behavior. In classical mechanics, the main cause
arising of chaotic motion is the exponentially rapid divergence of nearby trajectories that
in quantum mechanics  becomes impossible by reason of the Heisenberg uncertainty relation
\cite{Schust}. It is obvious, that in a standard representation of the quantum mechanics,
by analogy with the classical mechanics it is impossible to make  enter the criteria of the
exponentially rapid divergence of close wave functions since as already mentioned they are
deterministic functions. Moreover as it can be shown the analogue of Arnold's theorem on quantum
mapping,  different from the zero Planck's constant promotes to the suppression of chaos
\cite{Hannay}. In case when  $n$-dimensional classical dynamical system in the phase
space exhibits chaotic behavior and the sizes of the chaotic regions in the phase
space is lesser than $\hbar^n$, then its quantum analogue "does not see" such regions and,
respectively commits regular movement. In any case, the problem arises in the limit of
the classical motion when $\hbar\to0$, and respectively any classical system in the result of
such transition, by reason of Arnold's theorem necessarily becomes integrable, that
generally speaking is incorrect. Finally, the description of a quantum system is
seriously complicated, when sizes of regions of the classical chaotic motion become
larger, than the volume of the quantum cell $\hbar^n$.  In this case obviously chaos
should be appear also in the motion of the quantum system, i.e. should be chaotic the
wave function.

In recent years many studies on the problem of the quantum chaos  in systems with autonomous
and nonautonomous Hamiltonians have been conducted (see for example
\cite{CasCh,Schust,Gut,Hos,Blum,Casat,van}), however, the above problems as well
as a number  other issues regarding to the foundations of quantum mechanics, still
remain unresolved.

As well-known the general three-body problem is a typical example of a dynamic
system where on a large regions of the phase space are observed all features
of a complex motion including the bifurcation and chaos. In this paper, we
will consider  the problem of multichannel scattering in the classical three-body
system taking into account  the random external factors, in particular quantum
fluctuations, without using perturbation methods.
It should be noted that such consideration of the problem is very interesting,
from point of view its wide applications in applied problems, for example
at simulation of elementary chemical reactions taking into account randomness
of an environment etc.  Also it is important for a deeper understanding the
foundations of quantum mechanics, namely the principle of Bohr's correspondence
between the quantum system and, with its  non-integrable classical analog.

Note that the general three-body classical problem concerns the question of
understanding motions of three arbitrary point masses traveling
in space according to Newton's laws of mechanics. Many works on  celestial
and analytical mechanics, stellar and molecular dynamics,  devoted to the
study of this problem  are, as a rule, carried out by numerical simulation
(see for example \cite{Marchal,Bruno,Poinc,Arnold,Gutzwiller}).

Recently the author proved \cite{gev}, that the general classical three-body problem
may be reduced to the system of sixth order on the hypersurface of
the energy, which has the conform-Euclidean metric. The first three equations
of this system form the closed system of nonlinear partial differential equations
of the first order (the system of the Riccati's type equations). The main idea of
the work is that  the equations become random due to random exposure of the environment.
Mathematically, this is equivalent to assuming that the metric of space is random, and
respectively a motion of three-body system is described by the system of the stochastic
differential equations (SDEs) of Langevin type.
`

Finally, using the system SDEs, we derive the evolution equation describing the quantum
currents of the scattering process in the momentum representation, which is  solved in
combination with the system of classical equations of three-body.
In the work the criteria for the occurrence of quantum chaos is formulated.

\section{The classical three-body system}

The classic three-body problem in a most general formulation, is the problem of
multichannel scattering,  with series of possible asymptotic outcomes.  Schematically,
the scattering process can be represented as:
$$
1\,\,+\,(23)\quad\longrightarrow  \quad \begin{cases} 1\,\,+\,(23),
\\ 1+2+3,\\(12)\,\,+\,3,\\(13)\,\,+\,2,\\
\quad (123)^\ast\quad
\longrightarrow\quad
\begin{cases}1\,\,+\,(23),
\\ 1+2+3,\\(12)\,\,+\,3,\\(13)\,\,+\,
2,\\
(123)^{\ast\ast}\to\begin{cases}...\end{cases},
\end{cases}
\end{cases}
$$
where 1,2 and 3 are the separate particles, the bracket $(.)$ denotes bound states,
while $^{"\ast"}$ and $^{"\ast\ast"}$  denote some transition states of three-body system.

The aim of this study is the obtaining the equation describing the quantum probabilistic
currents going  between  asymptotic states.

The classical Hamiltonian of three-body system after Jacobi  and mass-scale  \cite{Delves1}
transformations can be written as (see also \cite{AshGev}):
\begin{equation}
\mathrm{H}(\textbf{r};\textbf{p})=\frac{\textbf{p}^2}{2\mu_0}
+\mathrm{V}(\textbf{r}), \label{01}
\end{equation}
where $\textbf{r}={\textbf{\emph{r}}}\oplus{\textbf{\emph{R}}}\in
\mathbb{R}^6$ and $\textbf{p}\in \mathbb{R}^6$ are correspondingly the position vector
and the momentum of the effective mass (imaginary point):
$$\mu_0=[m_1m_2m_3/(m_1+m_2+m_3)]^{1/2}.$$
Recall that $m_1,m_2$ and $m_3$ are masses of corresponding bodies).
With respect of the coordinate ${\textbf{\emph{r}}}$, that it represents the distance
between 2 and 3 particles, while the coordinate ${\textbf{\emph{R}}}$ designates the
distance between the particle 1 and the center of mass of  the  pair  (2,3)  (see Fig. 1), in
addition the total potential $\mathrm{V}(\textbf{r})$ depends from distances between partcles,
that  means that the interaction potential in fact depends from three variables.

Let us consider the following system of coordinates:
\begin{eqnarray}
\rho_1=r=||\textbf{\emph{r}}||,\qquad\rho_2=R=||\textbf{\emph{R}}||,\quad
 \rho_3=\theta,\quad \rho_4=\Theta,\quad \rho_5=\Phi,\quad
\rho_6=\Psi, \label{02}
\end{eqnarray}
where the first set of three coordinates;
$\{\bar{\rho}\}=(\rho_1,\rho_2,\rho_3)$ describes a position of an
imaginary point on the plane formed by bodies triangle (internal coordinates), while;
$\Theta\in(-\pi,+\pi]$, $\Phi=(-\pi,+\pi]$ and $\Psi\in[0,\pi]$ are
Euler angles describing rotation of the plane in $3D$ space (external coordinates). It is
clear that the full potential $\mathrm{V}(\textbf{r})$ in the new coordinates will depend
only on internal coordinates $\{\bar{\rho}\}$.
\begin{figure}
\center
\includegraphics[height=65mm,width=65mm]{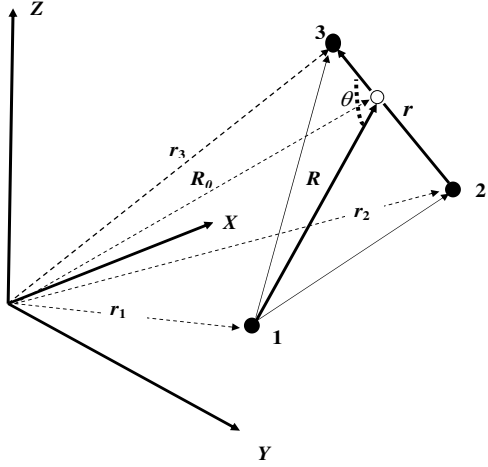}
 \caption{\emph{The Cartesian
coordinate system where the set of vectors}
$\textbf{\emph{r}}_1,\textbf{\emph{r}}_2$ and $\textbf{\emph{r}}_3$
\emph{denotes coordinates of the} 1, 2 \emph{and} 3 \emph{particles, respectively.
The  circle} $"\circ"$ \emph{denotes the center-of-mass of pair }(12)
\emph{which in the Cartesian system is expressed by}
$\textbf{\emph{R}}_0$. \emph{The Jacobi coordinates system described
by the radius-vectors }$\textbf{\emph{\emph{R}}}$ \emph{and
}$\textbf{\emph{r}}$, \emph{in addition to} $\theta$, \emph{denote
scattering angle.}} \label{fig1}
\end{figure}

The kinetic energy of three-body system in these variables has the form
(see also \cite{Fiz}):
\begin{eqnarray}
T=\frac{1}{2\mu_0}
\bigl\{\dot{{\textbf{\emph{r}}}}^2+\dot{{\textbf{\emph{R}}}}^2\bigr\}=\frac{1}{2\mu_0}
\Bigl\{\dot{{\emph{r}}}^{2}+
{{\emph{r}}}^{2}\bigl[\bm\omega\times\textbf{k}\bigr]^2+
\bigl(\dot{{\textbf{\emph{R}}}}+\bigl[\bm\omega\times
{\textbf{\emph{R}}}\bigr]\bigr)^2\Bigr\}, \label{16.a}
\end{eqnarray}
where  $\{\varrho\}$ the direction of unit vector $\textbf{k}$ in
the moving reference frame is defined by expression;
${\emph{\textbf{R}}}{\emph{\textbf{R}}}^{-1}=\pm \textbf{k}$. Below
the vector $\textbf{k}=(0,0,1)$ is directed towards a positive
direction of the axis $0Z$, while the angular velocity $\bm\omega$
describes rotation of the frame $\{\varrho\}$ with respect to
laboratory system. Having done a simple calculations in the
expression (\ref{16.a}) we can find:
\begin{equation}
T=\frac{1}{2\mu_0}\Bigl\{\dot{{{\emph{r}}}}^{\,2}+\dot{{{\emph{R}}}}^{2}+
{{{\emph{R}}}}^{2}\dot{\theta}^{\,2}+A{{{\emph{r}}}}^{2}+
B{{{\emph{R}}}}^{2}\Bigr\}, \label{17a}
\end{equation}
where  the following designations are made:
 $$
A=\bigl(\dot{\Theta}^2+\dot{\Phi}^2\sin^2\Theta\bigr)=\omega_X^2+\omega_Y^2,
\quad
 B=\bigl(\omega_X\cos\theta-\omega_Z\sin\theta\bigr)^2.
$$
Let us  note that at deriving  the expression (\ref{17a}) we have used
definition of the moving system $\{\varrho\}$ by the requirement
that unit vector $\bm\gamma$ lies in the plane $OXZ$ at the angle
$\theta$ to $OZ$, i.e. $\bm\gamma=(\sin\theta,0,\cos\theta)$. As
regards of projections of an angular velocity they satisfy
equations:
\begin{eqnarray}
\omega_X=\dot{\Phi}\sin\Theta\sin\Psi+\dot{\Theta}\cos\Psi,\nonumber\\
\omega_Y=\dot{\Phi}\sin\Theta\cos\Psi-\dot{\Theta}\sin\Psi,\nonumber\\
\omega_Z=\dot{\Phi}\cos\Theta-\dot{\Psi}. \label{18a}
\end{eqnarray}
Taking into account (\ref{17a}) and (\ref{18a}) we can find the
metric tensor:
$$
\gamma^{\alpha\beta}=\left(
  \begin{array}{cccccc}
    \gamma^{11}\, &\,  0\, & 0 \, & 0\,  & \, 0\, &\,  0 \\
    0\, & \gamma^{22} &\,  0 \, & 0\,  &\,  0\, &\,  0 \\
    0\, & \,0\, &  \gamma^{33} & 0\, & \,0 \,&\, 0 \\
    0\,  &\,  0\,  & \, 0\,  &\gamma^{44} &  \gamma^{45} &  \gamma^{46} \\
    0\,  & \, 0\, &\,  0\,  & \gamma^{54} & \gamma^{55} &  \gamma^{56}\\
    0\, &\, 0\,  &\,  0\, & \gamma^{64} &  \gamma^{65} &  \gamma^{66} \\
\end{array}
\right)
$$
\begin{equation}
 \label{19a}
\end{equation}
where the following designations are made:
$$
\gamma^{11}=\gamma^{22}=1,\quad\gamma^{33}={{\emph{R}}}^2,\quad
\gamma^{44}={r}^2+{R}^2\cos^2\Psi\cos^2\theta,
$$
$$
\gamma^{55}={r}^2\sin\Theta+{R}^2\bigl(\sin^2\Theta\sin^2\Psi\cos^2\theta+
\cos^2\Theta\sin^2\theta-2^{-1}\sin2\Theta\sin2\theta\sin\Psi\bigr)
$$
$$
\gamma^{66}={R}^2\sin^2\theta,\quad \gamma^{45}=\gamma^{54}=
{R}^2\bigl(\sin\Theta\sin2\Psi\cos^2\theta-2\cos\Theta\cos\Psi\sin2\theta\bigr),
$$
$$
\gamma^{46}=\gamma^{64}={R}^2\sin2\theta\cos\Psi,\quad
\gamma^{56}=\gamma^{65}= {R}^2\bigl(\sin\Theta\sin\Psi\sin2\theta-2\cos\Theta\sin^2\theta\bigr).
$$

Without going into details let us note that the considered problem has  12
integrals of motion using which the initial 18th order  system  is
reduced  to the 8th order system \cite{Whitt}.

\section{Classical three-body problem as the problem of geodesic flows on
 energy hypersurface}

As it is easy to see, the classical system of three bodies at motion
in the $3D$ Euclidean space permanently forms the triangle, and
hence Newton's equations describe the dynamical system on the space
of such triangles \cite{Fiz}. The last means that we can formally
consider the motion of a body-system consisting of two parts. The
first is the rotational motion of the body-triangle in the $3D$
Euclidian space and the second is the internal motion of bodies on
the plane defined by the triangle. Mathematically, the configuration
manifold of solid body ${R}^6$ can be represented as a direct
product of two subspaces \cite{Arnold1}:
$$ \mathbb{R}^6:\Leftrightarrow {\mathbb{R}}^3\times S^3,$$
where ${\mathbb{R}}^3$ is the manifold which is defined as an orthonormal
space of relative distances between  bodies  while $S^3$ denotes the
space of the rotation group $SO(3)$. However in the considered
problem, connections between bodies are not holonomic and
respectively we must change the representation for the configuration
manifold $\mathbf{M}:\Leftrightarrow {\mathbb{R}}^6$.

Let us consider the region of localization of  dynamical system
(further named \emph{the internal space} $\mathcal{M}_ t$):
\begin{eqnarray}
x^1=||{\textbf{\emph{r}}}||,\quad  x^2=
||{\textbf{\emph{R}}}||,\quad x^3=||{\textbf{\emph{r}}}+{\textbf{\emph{R}}}||
=\sqrt{(x^1)^2-2x^1x^2\cos\theta+(x^2)^2},
 \label{19b}
\end{eqnarray}
 where  $\theta\in[0,\pi]$ is the angle between the vectors
${\textbf{\emph{r}}}\in \bigl[0,\infty\bigr)$  and ${\textbf{\emph{R}}}\in \bigl[0,\infty\bigr)$
is the scattering angle in Jacobe coordinate system. The set of \emph{internal coordinates}
$\{\bar{x}\}=(x^1,x^2,x^3)\in \mathcal{M}_t$. The rotation of a
plane defined by  body-triangle will be described by the set of
three \emph{external coordinates} $(x^4,x^5,x^6)\in S^3_t$, where $
S^3_t$ is a space of the rotation group $SO(3)$ in a neighborhood of
interior points $M_i\{(x^1,x^2,x^3)_i\}\in\mathcal{M}_t$.

The subset
of all interior points  $\breve{\mathbf{M}}\subset\mathbf{M}$ is
represented as:
$$\breve{\mathbf{M}}\cong \mathcal{M}_t \times  S^3_t.$$
The set $\mathbf{M}\setminus \breve{\mathbf{M}}$ has zero measure
however in some cases it can be important for dynamics of the
classical three-body system.

 So, we can define a local system of coordinates in which further
 studies will be carried:
\begin{equation}
\overline{x^1,x^6}=\{x\}\in \breve{\mathbf{M}}.
\label{03}
\end{equation}
Taking into account the  Krylov's   well-known work  \cite{Krylov}, we will study the
motion of three-body system on the hypersurface of potential energy (HPE) of bodies system.
Note that the  HPE  is the curved space the  metric tensor of which is defined as follows:
\begin{eqnarray}
g_{\mu\nu}(\{x\})= g(\{x\})\delta_{\mu\nu}, \quad
g^{\mu\nu}=\delta^{\mu\nu}/g(\{x\}),\quad
g(\{x\})=\bigl[E-U(\{x\})\bigr]U_0^{-1}>0, \label{04}
\end{eqnarray}
where $E$ and $U(\{x\})\equiv{V}(\textbf{r})$ are  the total energy
and total interaction potential of bodies system  respectively,
$\delta_{\mu\nu}$  is the Kronecker symbol and
$U_0=\max||U(\{x\})||$ denotes maximal depth of the potential. In the case
the total potential depends on the relative distances between the particles;
$V(\textbf{r}_1,\textbf{r}_2,\textbf{r}_3)=V(|\textbf{r}_1-\textbf{r}_2|,|\textbf{r}_1-
\textbf{r}_3|,|\textbf{r}_2-\textbf{r}_3|)$, then the metric tensor can be written
in the internal space as; $g_{\mu\nu}(\{x\})=g_{\mu\nu}(\{\bar{x}\})$.

Now using the variational principle of Maupertuis we can derive
geodesic equations \cite{Arnold1,BubrNovFom}:
\begin{eqnarray}
\ddot{x}^{\alpha}+\Gamma^\alpha_{\beta\gamma} \dot{x}^{\,\beta}
\dot{x}^{\,\gamma} =0,\qquad \alpha,\beta,\gamma=\overline{1,6},
\label{05}
\end{eqnarray}
where $\dot{x}^\alpha=dx^\alpha/ds$ and $
\ddot{x}^{\,\alpha}=d^{\,2}x^\alpha/ds^2$; in addition  $s$ is a
scalar parameter of motion (e.g. the proper time), the  Christoffel
symbol; $ \Gamma^\alpha_{\beta\gamma}(\{x\})=\frac{1}{2}g^{\alpha
\mu}\bigl({\partial_\gamma g_{\mu \beta}}+\partial_\beta g_{\gamma
\mu}-\partial_\mu g_{\beta \gamma}\bigr),
 $
where $\partial_\alpha\equiv\partial_{x^\alpha}.$

Taking into account  the definition  for the metric tensor
(\ref{04})  from (\ref{05})  we can find the following system of
equations  describing geodesic flows on the potential energy
hypersurface:
\begin{eqnarray}
\ddot{x}^{1} =a_1\Bigl\{(\dot{x}^{1})^2-\sum_{\mu\neq
1,\,\mu=2}^6(\dot{x}^{\mu})^2\Bigr\}
+2\dot{x}^{1}\Bigl\{a_2\dot{x}^{2}+a_3
\dot{x}^{3} \Bigr\},
\nonumber\\
\ddot{x}^{2} =a_2\Bigl\{(\dot{x}^{\,2})^2-\sum_{\mu=1,\, \mu\neq
2}^6(\dot{x}^\mu)^2\Bigr\}
+2\dot{x}^{2}\Bigl\{a_3\dot{x}^{3}+a_1\dot{x}^{1} \Bigr\},
\nonumber\\
\ddot{x}^{3}
=a_3\Bigl\{(\dot{x}^{3})^2-\sum_{\mu=1,\,\mu\neq3}^6(\dot{x}^{\mu})^{2}\Bigr\}
+2\dot{x}^{3}\Bigl\{a_1 \dot{x}^{1}+a_2\dot{x}^{2} \Bigr\},
\nonumber\\
\ddot{x}^{4} =\dot{x}^{4}\Bigl\{a_1
\dot{x}^{1}+a_2 \dot{x}^{2}+a_3\dot{x}^{3}\Bigr\},\qquad\qquad\qquad\qquad\quad\,\,\,
\nonumber\\
\ddot{x}^{5}=\dot{x}^{5}\Bigl\{a_1
\dot{x}^{1}+a_2 \dot{x}^{2}+a_3\dot{x}^{3}\Bigr\},\qquad\qquad\qquad\qquad\quad\,\,\,
\nonumber\\
\ddot{x}^{6}=\dot{x}^{6}\Bigl\{a_1 \dot{x}^{1}+a_2
\dot{x}^{2}+a_3\dot{x}^{3}\Bigr\},\qquad\qquad\qquad\qquad\quad\,\,\,
\nonumber\\
\label{06}
\end{eqnarray}
where $ g(\{\bar{x}\})=g_{11}(\{\bar{x}\})=...=g_{66}(\{\bar{x}\})$ since the metric is the conformally Euclidean,
in addition; $a_i(\{\bar{x}\})=-(1/2)\partial_{x^i}\ln
g(\{\bar{x}\})$. \\
 In the system (\ref{06}), the last three equations
are integrated exactly:
\begin{equation}
\dot{x}^{\mu}=J_{\mu-3}/g(\{\bar{x}\}),\quad
J_{\mu-3}=const_{\mu-3},\label{07}
\end{equation}
where $\mu=\overline{4,6 }.$

 Note that $J_1,J_2$ and $J_3$ are integrals of motion.
They can be interpreted as projections of the total angular momentum
of three-body system $J=\sqrt{J^2_1+J^2_2+J^2_3}=const$ on
corresponding axis.

Substituting (\ref{07}) into equations (\ref{06}), we obtain the following
system of  non-linear second-order ordinary differential equations:
\begin{eqnarray}
\ddot{x}^1 =a_1\bigl\{(\dot{x}^1)^2-(\dot{x}^{2})^2-(\dot{x}^{3})^2-
\Lambda^2\bigr\}+ 2\dot{x}^1\bigl\{a_2\dot{x}^{2}+a_3
\dot{x}^{3} \bigr\},
\nonumber\\
\ddot{x}^2 =a_2\bigl\{(\dot{x}^2)^2-(\dot{x}^3)^2-(\dot{x}^1)^2-
\Lambda^2\bigr\}+ 2\dot{x}^{2}\Bigl\{a_3 \dot{x}^{3}+a_1
\dot{x}^{1}\bigr\},
\nonumber\\
\ddot{x}^{3} =a_3
\bigl\{(\dot{x}^{3})^2-(\dot{x}^{1})^2-(\dot{x}^{2})^2-
\Lambda^2\bigr\}+2\dot{x}^{3}\bigl\{a_1 \dot{x}^{1}+a_2\dot{x}^{2}\bigr\}, \label{08}
\end{eqnarray}
where $\Lambda(\{\bar{x}\})=(J/g\{\bar{x}\})^2.$

Doing designations;
\begin{equation}
\xi^1=\dot{x}^1,\quad \xi^2=\dot{x}^2, \quad   \xi^3=\dot{x}^3,
\label{07a}
\end{equation}
from the system (\ref{08}) it is possible to obtain the following system of non-linear first order
ODEs type of Ricatti:
\begin{eqnarray}
\dot{\xi^1} =a_1\bigl\{(\xi^1)^2-(\xi^{2})^2-(\xi^{3})^2-
\Lambda^2\bigr\}+ 2\xi^1\bigl\{a_2\xi^{2}+a_3
\xi^{3} \bigr\},
\nonumber\\
\dot{\xi}^2 =a_2\bigl\{(\xi^2)^2-(\xi^3)^2-(\xi^1)^2-
\Lambda^2\bigr\}+ 2\xi^{2}\Bigl\{a_3 \xi^{3}+a_1
\xi^{1}\bigr\},
\nonumber\\
\dot{\xi}^{3} =a_3
\bigl\{(\xi^{3})^2-(\xi^{1})^2-(\xi^{2})^2-
\Lambda^2\bigr\}+2\xi^{3}\bigl\{a_1 \xi^{1}+a_2\xi^{2}\bigr\}. \label{08a}
\end{eqnarray}

Thus, the system of equations (\ref{08}) or the six order ODEs system (\ref{07a})-(\ref{08a})
 describes the dynamics of an imaginary
point with an effective mass $ \mu_0 $, which moves on a Riemannian manifold;
 $\mathcal{M}=\bigl[\{\bar{x}\}\equiv(x^1,x^2,x^3)\in
\mathcal{M}_t;\,g_{ij}=\bigl(E-U(\{\bar{x}\})\bigr)U_0^{-1}\delta_{ij}>
0\bigr]$.

Finally with regard to  (\ref{04}) and  (\ref{07}) we can get the
reduced Hamiltonian:
\begin{eqnarray}
\mathcal{H}\bigl(\{\bar{x}\};\{\dot{\bar{x}}\}\bigr)=
g_{\mu\nu}(\{\bar{x}\})p^{\,\mu} p^{\,\nu}=
\frac{\mu_0}{2}g(\{\bar{x}\})\Bigl\{\sum_{i=1}^3\bigl(\dot{x}^{i}\bigr)^2+
\bigl[{J}/{g(\bar{x})}\bigr]^2\Bigr\}. \label{09}
\end{eqnarray}
Substituting (\ref{09}) into Hamilton equations:
\begin{equation}
\dot{x}^\mu=\frac{\partial \mathcal{H}}{\partial p^{\,\mu}},\qquad
\dot{p}^{\,\mu}=-\frac{\partial \mathcal{H}}{\partial x^\mu},
\label{10}
\end{equation}
 and by making simple calculations we can
get the system of geodesic equations (\ref{08}).

\section{The conditions of transformations of 6$D$ Euclidean space to the 6$D$ conformal-Euclidean space
}
At obtaining of equations system (\ref{08}), we have used some
physical considerations that from the mathematical point of view are
insufficiently rigorous,  to argue that the dynamical system
(\ref{08})  is equivalent to Newtonian problem of three-body. For a
strict proof of equivalence of approaches, we need to prove that
there is one-to-one mapping between two sets of coordinates;
$\overline{\rho^1,\rho^6}=\{\rho\}$ and $\overline{x^1,x^6}=\{x\}$.

Let us consider two spaces $\mathbf{E}^6\cong \mathbb{R}^6$ and $ \mathbf{M}$
which satisfy to the condition of one-to-one mapping;
$\mathbf{E}^6:\Leftrightarrow \mathbf{M}$. We will suppose that the
Euclidean space $\mathbf{E}^6$ is defined by the set of coordinates
$\{\rho\}$ and the metric tensor $\gamma_{\mu\nu}(\{\rho\})$, while  the
frame;  $\{x\}$ and the metric tensor
$g_{\mu\nu}(\{x\})$ respectively. The linear infinitesimal element
in  both coordinate systems can be represented as:
\begin{eqnarray}
(ds)^2=
\gamma^{\alpha\beta}(\{\mathrm{\rho}\})d\mathrm{\rho}_{\alpha}
d\mathrm{\rho}_\beta=
g_{\mu\nu}(\{x\})dx^\mu{dx^\nu},\quad
\alpha,\beta,\mu,\nu=\overline{1,6}, \quad
\label{11}
\end{eqnarray}
where the metric tensor $g_{\mu\nu}(\{x\})$ is defined as follows:
\begin{equation}
g_{\mu\nu}(\{x\})=\gamma^{\alpha\beta}(\{\mathrm{\rho}\})\rho_{\alpha;\mu}\rho_{\beta;\nu},
 \label{12}
\end{equation}
where $\mathrm{\rho}_{\alpha;\mu}=\partial\mathrm{\rho}_\alpha/\partial{x}^\mu.$

Since tensor $g_{\mu\nu}(\{x\})$ is still defined in a rather
arbitrary way we can impose  additional conditions on it. In
particular we will require that the metric tensor $g_{\mu\nu}({x})$
 describes the conformal-Euclidean space;
$g_{\mu\nu}({x})=g(\{x\})\delta_{\mu\nu}$ (see (\ref{04})). The last
means that the following algebraic equations must be satisfied:
\begin{eqnarray}
\gamma^{\alpha\beta}(\{\rho\})
\mathrm{\rho}_{\alpha;\mu}\mathrm{\rho}_{\beta;\nu}=g(\{x\})\delta_{\mu\nu}.
\label{13}
\end{eqnarray}

As one can sure, the system of algebraic equations (\ref{13}) is
underdetermined since it consists from 21 equations while the number
of unknown variables is 36. It is obvious that these equations
are compatible, then the system (\ref{13}) has an infinite number of
real and complex solutions. These solutions form two different
manifolds of 15th order. For a classical problem real solutions are
the important ones, therefore below we will investigate properties of a
real manifold. In a similar way we can obtain the system of
algebraic equations for inverse transformations:
\begin{eqnarray}
\gamma_{\alpha\beta}(\{\rho\})g^{-1}(\{\bar{x}\})=
x^{\mu}_{;\,\alpha}x^{\nu}_{;\,\beta}\delta_{\mu\nu}, \label{14}
\end{eqnarray}
where $x^{\mu}_{;\,\alpha}=\partial x^\mu/\partial{\rho}^\alpha$.

It is obvious that if there are direct transformations then there
are inverse transformations too.

Let us make  new designations:
\begin{eqnarray}
\mathrm{x}_\mu=\rho_{1;\mu},\quad \mathrm{y}_\mu=\rho_{2;\mu}, \quad
\mathrm{z}_\mu=\rho_{3;\mu}, \quad \mathrm{u}_\mu=\rho_{4;\mu},\quad
\mathrm{v}_\mu=\rho_{5;\mu},\quad \mathrm{w}_\mu=\rho_{6;\mu}.
\label{20}
\end{eqnarray}
 Taking into account the fact that the tensor; $g_{\mu\nu}(\{\bar{x}\})$
still is an arbitrary one, we can require fulfillment of  following
conditions for its elements:
\begin{eqnarray}
\mathrm{x}_4=\mathrm{x}_5=\mathrm{x}_6=0,\,\quad \mathrm{y}_4=\mathrm{y}_5=\mathrm{y}_6=0,\,\quad
\nonumber\\
\mathrm{z}_4=\mathrm{z}_5=\mathrm{z}_6=0,\,
\quad\,
\mathrm{u}_1=\mathrm{u}_2=\mathrm{u}_3=0,\quad
\nonumber\\
\mathrm{v}_1=\mathrm{v}_2=\mathrm{v}_3=0,\quad
\mathrm{w}_1=\mathrm{w}_2=\mathrm{w}_3=0. \quad
 \label{21}
\end{eqnarray}
 Using (\ref{19a}), (\ref{20}) and conditions (\ref{21}) from the equation (\ref{13})  we can
  obtain two independent  systems of algebraic equations:
\begin{eqnarray}
\mathrm{x}_1^2+\mathrm{y}_1^2+\gamma^{33}\mathrm{z}_1^2\,=\,g(\{\bar{x}\}),
\nonumber\\
\mathrm{x}_2^2+\mathrm{y}_2^2+\gamma^{33}\mathrm{z}_2^2\,=\,g(\{\bar{x}\}),
\nonumber\\
\mathrm{x}_3^2+\mathrm{y}_3^2+\gamma^{33}\mathrm{z}_3^2\,=\,g(\{\bar{x}\}),
\nonumber\\
\mathrm{x}_1\mathrm{x}_2+\mathrm{y}_1\mathrm{y}_2+\gamma^{33}\mathrm{z}_1\mathrm{z}_2=0,
\nonumber\\
\mathrm{x}_1\mathrm{x}_3+\mathrm{y}_1\mathrm{y}_3+\gamma^{33}\mathrm{z}_1\mathrm{z}_3=0,
\nonumber\\
\mathrm{x}_2\mathrm{x}_3+\mathrm{y}_2\mathrm{y}_3+\gamma^{33}\mathrm{z}_2\mathrm{z}_3=0,
\label{22}
\end{eqnarray}
and correspondingly:
\begin{eqnarray}
\gamma^{44}\mathrm{u}_4^2+
\gamma^{55}\mathrm{v}_4^2+\gamma^{66}\mathrm{w}_4^2+2(\gamma^{45}\mathrm{u}_4\mathrm{v}_4
+\gamma^{46}\mathrm{u}_4\mathrm{w}_4+\gamma^{56}\mathrm{v}_4\mathrm{w}_4)=g(\{\bar{x}\}),
\nonumber\\
\gamma^{44}\mathrm{u}_5^2+
\gamma^{55}\mathrm{v}_5^2+\gamma^{66}\mathrm{w}_5^2+2(\gamma^{45}\mathrm{u}_5\mathrm{v}_5
+\gamma^{46}\mathrm{u}_5\mathrm{w}_5+\gamma^{56}\mathrm{v}_5\mathrm{w}_5)=g(\{\bar{x}\}),
\nonumber\\
\gamma^{44}\mathrm{u}_6^2+
\gamma^{55}\mathrm{v}_6^2+\gamma^{66}\mathrm{w}_6^2+2(\gamma^{45}\mathrm{u}_6\mathrm{v}_6
+\gamma^{46}\mathrm{u}_6\mathrm{w}_6+\gamma^{56}\mathrm{v}_6\mathrm{w}_6)=g(\{\bar{x}\}),
\nonumber\\
a_4\mathrm{u}_4+a_5\mathrm{v}_4+a_6\mathrm{w}_4=0,
\nonumber\\
b_4\mathrm{u}_5+b_5\mathrm{v}_5+b_6\mathrm{w}_5=0,
\nonumber\\
c_4\mathrm{u}_6+c_5\mathrm{v}_6+c_6\mathrm{w}_6=0.
 \label{23}
\end{eqnarray}
In equations (\ref{23}) the following designations are made:
$$a_{i}=\gamma^{i4}\mathrm{u}_5+\gamma^{i5}\mathrm{v}_5+\gamma^{i6}\mathrm{w}_5,\quad
b_{j}=\gamma^{j4}\mathrm{u}_6+\gamma^{j5}\mathrm{v}_6+\gamma^{j6}\mathrm{w}_6, \quad
c_{k}=\gamma^{k4}\mathrm{u}_4+\gamma^{k5}\mathrm{v}_4+\gamma^{k6}\mathrm{w}_4,$$
where $i=\overline{4,6}.$

Note that solutions of algebraic systems (\ref{22}) and (\ref{23}) form two different
manifolds of 3rd order, that are in one-to-one mapping with the  potential energy
hypersurface of the three-body system. Recall that analogical systems of algebraic
equations can be obtained for inverse  transformations. It is easy to see that
coordinates which are  defined by formulas  (\ref{02}) and (\ref{19b})  in general
case do not satisfy to conditions of transformations (\ref{22}) and (\ref{23}) and
correspondingly the system of equations (\ref{08}) generally speaking is not equivalent to
the Newtonian three-body problem. However, we have proved that there exists a system of
local coordinates (\ref{03}) that satisfy to systems of algebraic equations
(\ref{22})-(\ref{23}) and correspondingly with consideration of this circumstances,
we can affirm that the equations system (\ref{08}) is equivalent to the three-body Newtonian
problem. Recall that in this case the equations (\ref{08}) conserve their previous form
and only the dependence of the potential from coordinates is changed.

The coordinate transformations between  two sets  of internal coordinates $\{\bar{\rho}\}$ and
$\{\bar{x}\}$ are easy to  represent in differential form:
\begin{eqnarray}
\rho_1=\rho_1^0+d\rho_1, \qquad d\rho_1=\mathrm{x}_1dx^1+\mathrm{x}_2dx^2+\mathrm{x}_3dx^3,
\nonumber\\
\rho_2=\rho_2^0+d\rho_2, \qquad d\rho_2=\mathrm{y}_1dx^1+\mathrm{y}_2dx^2+\mathrm{y}_3dx^3,
\nonumber\\
\rho_3=\rho_3^0+d\rho_3, \qquad d\rho_3=\mathrm{z}_1dx^1+\mathrm{z}_2dx^2+\mathrm{z}_3dx^3,\,\,
\label{24}
\end{eqnarray}
where $\{\bar{\rho}^0\}=(\rho_1^0,\rho_2^0,\rho_3^0)$ denotes the initial point.

Recall that at every stage of the evolution of a dynamical system the number of selection of
the local coordinates system, as expected is unlimited. This fact in particular is reflected
in infinite number of sets of solutions $[(\mathrm{x}_1,..),(\mathrm{y}_1,..),(\mathrm{z}_1,..)]_{i}$
of the algebraic system (\ref{22}).

\section{Classical movement under the influence of quantum fluctuations}

Let us assume that the dynamical system at the movement undergoes to the  influence  random forces,
in particular to quantum fluctuations. In a mathematical sense, it is equivalent to the fact that the
metric tensor and the corresponding coefficients in equations (\ref{08a}) are random functions:
$$Q_f:a_i(\{\bar{x}\})\mapsto \tilde{a}_i(\{\bar{x}(s)\})=\bar{a}_i(\{\bar{x}(s)\})+\eta_i(s), $$
and
$$Q_f:\Lambda^2(\{\bar{x}\})\mapsto \tilde{\Lambda}^2(\{\bar{x}(s)\})=\bar{\Lambda}^2(\{\bar{x}(s)\})+
\eta_0(s),$$
where $\bar{a}_i(\{\bar{x}(s)\})$ and $\bar{\Lambda}^2(\{\bar{x}(s)\})$ are regular functions,
the function $Q_f$ displays random influences, the set of functions $\{\eta_0(s),...,\eta_3(s)\}$
denote random generators which will be refined below. It is obvious that  the random component of $
\tilde{a}_i $  by a value is much more than a random member in the  $ \tilde{\Lambda} $,
since $\tilde{a}_i$ is the first derivative of metric tensor.

With consideration of said the system of equations (\ref{08a}) can be expanded and presented in
the form of stochastic equations of  Langevin type:
\begin{eqnarray}
\dot{\xi}^i=A^i(\{\bar{\xi}\}|\{\bar{x}(s)\})+\sum_{j=1}^3B^{ij}(\{\bar{\xi}\}|\{\bar{x}(s)\})\eta_j(s)+O(\eta^2),
\qquad i=\overline{1,3},
\label{25}
\end{eqnarray}
where $\{\bar{\xi}\}=(\xi^1,\xi^2,\xi^3),$ in addition:
$$
A^1(\{\bar{\xi}\}|\{\bar{x}(s)\})=\bar{a}_1\bigl\{(\xi^1)^2-(\xi^2)^2-(\xi^3)^2-
\bar{\Lambda}^2\bigr\}+2\xi^1(\bar{a}_2\xi^2+\bar{a}_3\xi^3),
$$
$$
A^2(\{\bar{\xi}\}|\{\bar{x}(s)\})=\bar{a}_2\bigl\{(\xi^2)^2-(\xi^1)^2-(\xi^3)^2
-\bar{\Lambda}^2\bigr\}+2\xi^2(\bar{a}_3\xi^3+\bar{a}_1\xi^1),
$$
$$
A^3(\{\bar{\xi}\}|\{\bar{x}(s)\})=\bar{a}_3\bigl\{(\xi^3)^2-(\xi^2)^2-(\xi^1)^2
-\bar{\Lambda}^2\bigr\}+2\xi^3(\bar{a}_1\xi^1+\bar{a}_2\xi^2),
$$
and, respectively;
$$
 B^{11}= (\xi^1)^2-(\xi^2)^2-(\xi^3)^2-
\bar{\Lambda}^2,\quad  B^{12}=2\xi^1\xi^2,\quad B^{13}=2\xi^1\xi^3,
$$
$$
B^{21}=2\xi^2\xi^1,\quad B^{22}=(\xi^2)^2-(\xi^1)^2-(\xi^3)^2
-\bar{\Lambda}^2,\quad B^{23}=2\xi^2\xi^3,
$$
$$
 B^{31}=2\xi^3\xi^1,\quad  B^{32}=2\xi^3\xi^2,\quad B^{33}=(\xi^3)^2-(\xi^2)^2-(\xi^1)^2
 -\bar{\Lambda}^2.
$$
Note, that $\{\bar{x}(s)\}$ is the external parameter and is defined by solution  the
classical equations' system (\ref{08}). In other words the variable $\xi_i$ depends from
$\{\bar{x}(s)\}$ parametrically and, accordingly it is correctly  written in the form $\xi^i(s|\{\bar{x}(s)\})$.
Recall that the cause of the stochastic motion of the system can be also random external force. At study
the molecular dynamics will be more natural and important if we consider the quantum fluctuations. In
this case the stochastic equations of motion (\ref{25}) can be written in the  form:
\begin{eqnarray}
\dot{\xi}^i=A^i(\{\bar{\xi}\}|\{\bar{x}(s)\})+\eta_i(s),
\qquad i=\overline{1,3},
\label{25a}
\end{eqnarray}
where, it is natural to put, $\eta(s)=\eta_1(s)=\eta_2(s)=\eta_3(s)$.

If to assume that the average value of stochastic force is zero $<\eta(s)>=0$,
then  averaging of the  stochastic equations (\ref{25a}) on relatively small scales
of interval $"s"$ leads to the system of the classical equations (\ref{08a}). Note
that the same we cannot say in respect to stochastic equations (\ref{25}), since
their  averaging, generally speaking does  not lead to  the classical equations (\ref{08a}).

The joint probability density for the independent variables can be formally represented as:
\begin{eqnarray}
P(\{\bar{\xi}\},s|\{\bar{x}(s)\})=\prod_{i=1}^3\bigl\langle\delta\bigl[\xi^i(s|\{\bar{x}(s)\})-\xi^{i}\bigr]\bigr\rangle.
\label{26}
\end{eqnarray}
Differentiating the expression (\ref{26}) by variable $"s"$ with considering the system
of stochastic equations (\ref{25}) and assuming that the stochastic generators satisfy
to correlation properties of the \emph{white noise}:
\begin{equation}
\langle\eta_i(s)\rangle=0, \qquad\langle\eta_i(s)\eta_j(s')\rangle=2\epsilon_{ij}\delta(s-s'),
\label{25ab}
\end{equation}
where the constant $\epsilon_{ij}$ describes the power of random fluctuations, for the joint
probability density the following equation may be obtained \cite{Lif}  (more detail see \cite{Kljat}):
\begin{eqnarray}
\frac{\partial P}{\partial s}=\sum_{i=1}^3\frac{\partial }{\partial \xi^i}\bigl(A^iP\bigr)+
\sum_{i,j,\,l,\,k=1}^3\epsilon_{ij}\frac{\partial }{\partial \xi^l}\Bigl[B^{il}
\frac{\partial }{\partial \xi^k}\bigl(B^{kj}P\bigr)\Bigr].
\label{25b}
\end{eqnarray}
In the case of the quantum fluctuations, the equation of joint probability density is
simplified (see equation (\ref{25a})), and it may be found from (\ref{25b}) if to put
$B^{ij}\equiv1$ and $\epsilon_{ij}=\delta_{ij}\epsilon$, where $\delta_{ij}$ denotes
the Kroneker symbol. Recall, that the power of the quantum fluctuations is defined by
the expression,  $\varepsilon=\hbar\sqrt{<\omega^2>}/2$, where $<\omega^2>$  is the
frequency dispersion of randomly fluctuating virtual fields.

Thus, the equation (\ref{25b}) with the system of equations (\ref{08}) and coordinate
transformations (\ref{24}) describes the classical multichannel scattering in the three-body
system taking into account the quantum fluctuations. Finally, note that the function $P(\{\bar{\xi}\},s|
\{\bar{x}(s)\})$ is a probability of finding the momentum of the point mass with mass
$\mu_0$ in the range $[\bm{\xi},\, \bm{\xi}+d\bm{\xi}]$ and therefore it can be interpreted
as $P=|\psi(\bm{\xi},s)|^2$, where $\psi(\bm{\xi},s)$ denotes the full wave function of
the three-body system in the momentum representation, while $\bm{\xi}\equiv \{\bar{\xi}\}$
is the $3D$ momentum  in the units $\mu_0$.

\section{Conclusion}
As it is well-known, the molecular dynamics in three-body system is often studied both in the
framework of classical mechanics as well as by methods of the quantum mechanics. This
is due to the fact that molecular systems depending on values initial parameters
(collision energy, mass of particles etc.) can demonstrate all diversity of classical
and quantum motions.  Despite the fact that the quantum
mechanics is a more general representation than the classical mechanics, nonetheless
as claimed by many researchers, often the classical calculations more accurately
describe the nature of molecular dynamics than  the quantum calculations.
This fact is not accidental, since the molecular systems as a rule in large
regions of phase space often demonstrate chaotic motion, which is not anti-aliased
by the quantum uncertainties, by given above reasons.

For solving the above stated problems, we made extension of the classical three-body problem,
assuming that the metric of the conform-Euclidean space has a random component.  In
result of this we obtained the system of Langevin type SDEs (\ref{25}) and also (\ref{25a}),
which describe dynamics the three-body system under the influence of random forces, origin
of which can be different, in particular can be quantum fluctuations.
Assuming, that the fluctuations satisfy the correlation conditions of the white noise and,
using SDEs (\ref{25}) we get the evolution equation for the quantum probability currents
in the form of a partial second order differential equation (\ref{25b}). Note that the equation
(\ref{25b}) can be solved combining with the classical equations of motion (\ref{08a}) and
algebraic equations (\ref{22})  taking into account coordinate transformations (\ref{24}).

Now obviously, if the solution of
the classical system has a chaotic behavior, then this will be reflected on properties
of quantum probability currents and correspondingly on the wavefunction of bodies system.
To assess the nature of the propagation of quantum probability currents arising at
the process of multichannel scattering in the three-body system, it is useful to
define the criterion of the discrepancy of two arbitrary probability flows.
Following to definition of the Kullback-Leibler  relatively  the distance between
two continuous distributions \cite{Kul},  $P_a=P(\{\bar{\xi}\},s|
\{\bar{x}_a(s)\})$ and $P_b=P_b(\{\bar{\xi}\},s|\{\bar{x}_b(s)\})$, we can determine
the distance between the two tubes probabilistic quantum currents in the following way:
$$
D_{ab}(s)=\int_{\mathcal{M}_t}P(\{\bar{\xi}\},s|
\{\bar{x}_a(s)\})\ln\frac{P(\{\bar{\xi}\},s|
\{\bar{x}_a(s)\})}{P(\{\bar{\xi}\},s|
\{\bar{x}_b(s)\})} d\xi^1d\xi^2d\xi^3,
$$
where it is assumed that  $\{\bar{x}_a(s)\in\mathcal{M}_t$ and
$\{\bar{x}_b(s)\}\in\mathcal{M}_t$ are two different trajectories.
If at some point in time $ s_0 $, these trajectories come out from the single point
$\{\bar{x}^0\}$, then we say that in this moment of time the probability distributions,
and respectively the  wave functions have zero distance or they are the same. The
trajectories as well as the corresponding distributions can diverge during the time. In case the distance
between two flows depending on time exponentially grow, i.e., $D_{ab}(s)\sim e^{ks}$,
where $k>0$ is some constant, then there is every reason to believe that the
quantum system wavefunction is chaotic and hence the system is quantum-chaotic.

Lastly important to note that the full quantum theory of multichannel scattering in
the three-body system allowing the emergence of quantum chaos in the wave function,
can be constructed based on Schr\"{o}dinger equation with the Hamiltonian (\ref{09}),
also taking into account  the classical equations (\ref{08a}), (\ref{22}) and (\ref{24}).
Recall that the system of the classical equations in this case is responsible for the
topological peculiarities of tubes of the quantum probabilistic currents and transitions
between asymptotic channels.

 \end{document}